\documentclass[12pt]{article}

\usepackage{amsmath, amssymb, amscd,epsfig,wrapfig}

\usepackage[english]{babel}
\usepackage[koi8-r]{inputenc}
\begin{document}

\begin{center}
{\Large\bf On the Possibility of Development of the Explosion Instability
in a Two-Component Gravitating System}
\end{center}
\ 
A. S. Kingsep, Yu. A. Fridman \\
\ 

\qquad\hrulefill\qquad\qquad
\\[6mm]

{\it We obtain an expression for the energy of the density wave 
propagating in a multicomponent gravitating medium in the 
form well known from electrodynamics.  Using the above, 
the possibility of "triple production" of the quasiparticles, 
or waves, with their energies summing up to zero, in 
a nonequilibrium medium is demonstrated.  That kind of 
resonance wave interaction is shown to result in the
development of an explosion instability.  By the method
developed in plasma physics, the characteristic time
of the instability is evaluated.}

\section{Introduction}

One would judge dark matter to be a natural object for
the collisionless hydrodynamics of multicomponent
gravitating systems, in which the particle movement is
governed by the self-consistent gravitational potential.
Up till now, its particles have been manifesting themselves
solely in gravitational interaction, hence the lack of the 
evidence on the question of their nature (see for instance
\cite{Gurevich}).  In spite of certain well-known achievements
of the last years (like a bona fide Hubble photograph of
2004, of the two clasters colliding, which is strongly 
believed to have captured a ring of dark matter), the
question of the very existence of the dark matter,
raised by Fritz Zwicky in 1937 (see \cite{Zwicky}),
is still unaswered \cite{Copperstock}.

Development of the instabilities in a multicomponent system
is determined by the hydrodynamical (or kinetic) parameters
of the media composing it.  Traces of structures or
turbulences owing their appearance to this or that 
instability would be impossible to observe in the dark
matter, but they should be present, say, in the gaseous
disk.  A list of instabilities capable of arising in the
system "gaseous disk --- dark matter" might come handy
when one sets to drawing some conclusions concerning
hydrodynamical (or kinetic) properties of the dark matter
as a medium.  Linear hydrodynamical beam-jeans
instabilities of a multicomponent gravitating medium
are described, for instance, in \cite{PF instabilities},
\cite{FF}.  If the media are sufficiently hot, the
kinetic beam-jeans instability can develop instead of
the hydrodynamical one.  If the system is stable in the
linear approximation (e. g., the large-scale instabilities
are stabilized by rotation \cite{PF gravitational
systems}), non-linear scenarios of the development of
instabilities enter the game.

\section{Energy of the density wave in a gravitating medium}

In this section, our aim is to obtain an expression 
for the energy of the density wave in a gravitating medium
in the form convenient for further investigations.

In \cite{LL Hydro}, while deriving the formula
expressing the energy of the density wave in an ordinary
gas (i. e., self-gravitation amounting to effective
zero), total energy of the unit volume of 
a homogeneous medium is given as follows:  $E = \rho 
\epsilon + \frac{\rho v^2}{2}$, $\rho$ being the
density of the medium, $\epsilon$ --- the mass density
of the internal density of the medium,  $v$ --- its
total velocity.  Taking self-gravitation into account,
one should add a term corresponding to the potential
energy $\frac{1}{2}\rho \psi$, where $\psi$ is the
gravitational potential.  Apart from that following
the standard procedure, for the density wave in the
motionless medium we get:
\begin{equation}
\label{energy one-component medium}
E_{wave} = \frac{1}{2}\frac{c_s^2}{\rho_0} \tilde \rho^2 +
\rho_0 \frac{\tilde v^2}{2} + \frac{1}{2}\tilde \rho \tilde \psi
\end{equation}
Here the "tilde" symbol marks the perturbed values while
the index $"0"$ stands for the equilibrium,  $c_s$ is the
speed of the sound in the gas (we consider the isotermic
case here, so that $c_s^2 = p/\rho$, $p$ being the gas
pressure).  In a frame of reference where the medium has
the velocity $v_0$, in the expression for the density
of the energy of the wave one more term would appear,
amounting to $\tilde \rho v_0 \tilde v$.

In the more general case of multicomponent gravitating 
medium with the plane-parallel relative motion of the
components, the energy of the density wave per unit
volume acquires the form:
\begin{equation}
\label{energy multi-component medium}
E_{wave} = \sum_i \left( \frac{\rho_{0i} \tilde v_{i}^2}{2} + \tilde \rho_i \mathbf{v_{0i}} \mathbf{\tilde v_i} + \frac{c_{si}^2 \tilde \rho_i^2}{2 \rho_{0i}} + \frac{1}{2}\tilde \rho_i \tilde \psi_i \right)
\end{equation}
where the index "i" denotes the (both perturbed and
pertaining to the equilibrium state) parameters of the
component of that same number.

The equations of gravihydrodynamics for the multicomponent
gravitating medium have the following form:
{\large
\begin{equation}
\label{hydrodynamics}
\left\{
\begin{array}{rcl}
\frac{\partial \rho_i}{\partial t} + div(\rho_i \mathbf{v_i}) &=& 0 \\
\frac{\partial \mathbf{v_i}}{\partial t} + (\mathbf{v_i} \nabla) \mathbf{v_i} &=& - \nabla \psi - \frac{\nabla p_i}{\rho_i} \\
\Delta \psi &=& 4 \pi G \sum_i \rho_i
\end{array}
\right.
\end{equation}}
"i" being any integer from $1$ to $n$, $n$ --- the number
of the components of the system under consideration.

Linearizing the equations and expanding them into Fourier
series, for any given harmonic $\varpropto e^{- i \omega t + i\mathbf k \mathbf r}$ we get
\begin{equation}
\label{density vs potential}
\tilde \rho_i = \frac{\rho_{0i} k^2 \tilde \psi}{\left(\omega - \mathbf k \mathbf{v_{0i}}\right)^2 - k^2 c_{si}^2}
\end{equation}

\begin{equation}
\mathbf{\tilde v_i} = \frac{\mathbf k \tilde \psi \left(\omega - \mathbf k \mathbf{v_{0i}}\right)}{\left(\omega - \mathbf k \mathbf{v_{0i}}\right)^2 - k^2 c_{si}^2}
\end{equation}

Substituting the above into the expression (\ref{energy multi-component medium})
for the energy of the density wave in a multicomponent medium,
we get:
\begin{equation}
\label{energy wave almost}
E_{wave} = k^2 \tilde \psi^2 \sum_i \frac{\rho_{0i}\omega \left(\omega - \mathbf k \mathbf{v_{0i}} \right) }{\left(\left(\omega - \mathbf k \mathbf{v_{0i}}\right)^2 - k^2 c_{si}^2\right)^2} 
\end{equation}

This is almost what we have aimed at.  

Now, the dispersion equation derived from the linearized
equations of gravihydrodynamics:
\begin{equation}
\label{dispersion equation}
\varepsilon_0 = 1 + \sum_i \frac{\omega_{ji}^2}{\left(\omega - \mathbf k \mathbf v_{0i} \right)^2 - k^2 c_{si}^2} = 0,
\end{equation}
$\omega_{ji}$ being the jeans frequency of the component number
$i$ of the medium, defined in the usual way by 
$\omega_{ji}^2 = 4 \pi G \rho_{0i}$, $G$ being the newtonian
gravitational constant, and the physical interpretation
of the value $\varepsilon_0$ will be discussed a few
passages below.  Direct calculations show that for the
values $\omega$ and $\mathbf k$ satisfying the dispersion
equation (\ref{dispersion equation}),
\begin{equation}
\label{wave energy plasma-like}
E_{wave} = \frac{\partial (\varepsilon_0 \omega)}{\partial
\omega} \left(- \frac{|\mathbf{\tilde g}|^2}{8 \pi G}
\right)
\end{equation}
where $\mathbf{\tilde g} = - \nabla \tilde \psi$ is the
perturbed gravitational field strength (for a given
Fourier harmonic $\mathbf{\tilde g} = - i k \tilde \psi$).

The structure of the expression (\ref{wave energy
plasma-like}) is fully analogous to the well-known
results from classical electrodynamics \cite{LL Electro}
for the energy of the electromagnetic wave or the wave
of the electric charge density in a dielectric medium.
The difference rests in the sign (and in the presence,
in the denominator, of the newtonian constant $G$), 
reflecting the difference in the nature of 
coulombian and newtonian interaction).

The value $\varepsilon_0$, analogous to the dielectric
permettivity of a medium to the electric charge density
waves, characterizes the responce of the density 
distribution in the various components of the gravitating
medium to the perturbation of the gravitational potential,
as it can be seen, for instance, from the expression
for the perturbed density (\ref{density vs potential}).

So we see that the sign of the energy of a density wave
in a multicomponent gravitating medium is determined
by the sign of {\Large$\frac{\partial
\varepsilon_0}{\partial \omega}$}.

\section{On the Possibility of Simultaneous Existence 
of the Waves of Positive and Negative Energy in a 
Multicomponent Medium}

Consider the simplest case of a multicomponent medium ---
a case of a medium consisting of two streams of matter.
Let both media composing the system have the same 
hydrodynamical parameters (the isothermic sound speed 
$c_s$ and jeans frequency $\omega_j^2 = 4 \pi G \rho$) 
and move through each other with the velocities
$v_0$ and $-v_0$.  For "one-dimensional" wave 
($\mathbf k \| \mathbf v_0$) \footnote{The case of
$\mathbf k \bot \mathbf v_0$, as can be easily shown,
corresponds to an ordinary jeans instability.
But the waves of large wavelength  $(k^2c_s^2 < 2\omega_j^2)$ 
are excluded from out consideration anyway --- by the limits
of validity of our plane approximation as well as by the
properties of real astronomical self-gravitating objects
and their rotational geometry, in which the large-scale
instabilities tend to be suppressed, see, for instance,
\cite{PF gravitational systems}} in such a medium, (\ref{dispersion equation})
takes the form:
\begin{equation}
\label{dispersion doublecomponent}
\varepsilon_0 = 1 + \frac{\omega_{j}^2}{\left(\omega - k v_0\right)^2 - k^2 c_{s}^2} + \frac{\omega_{j}^2}{\left(\omega + k v_0\right)^2 - k^2 c_{s}^2} = 0
\end{equation}

A sketch of behaviour of the left side of the dispersion 
equation (as dependent on frequency) is shown on Fig. 1.

\begin{center}
\epsfig{file=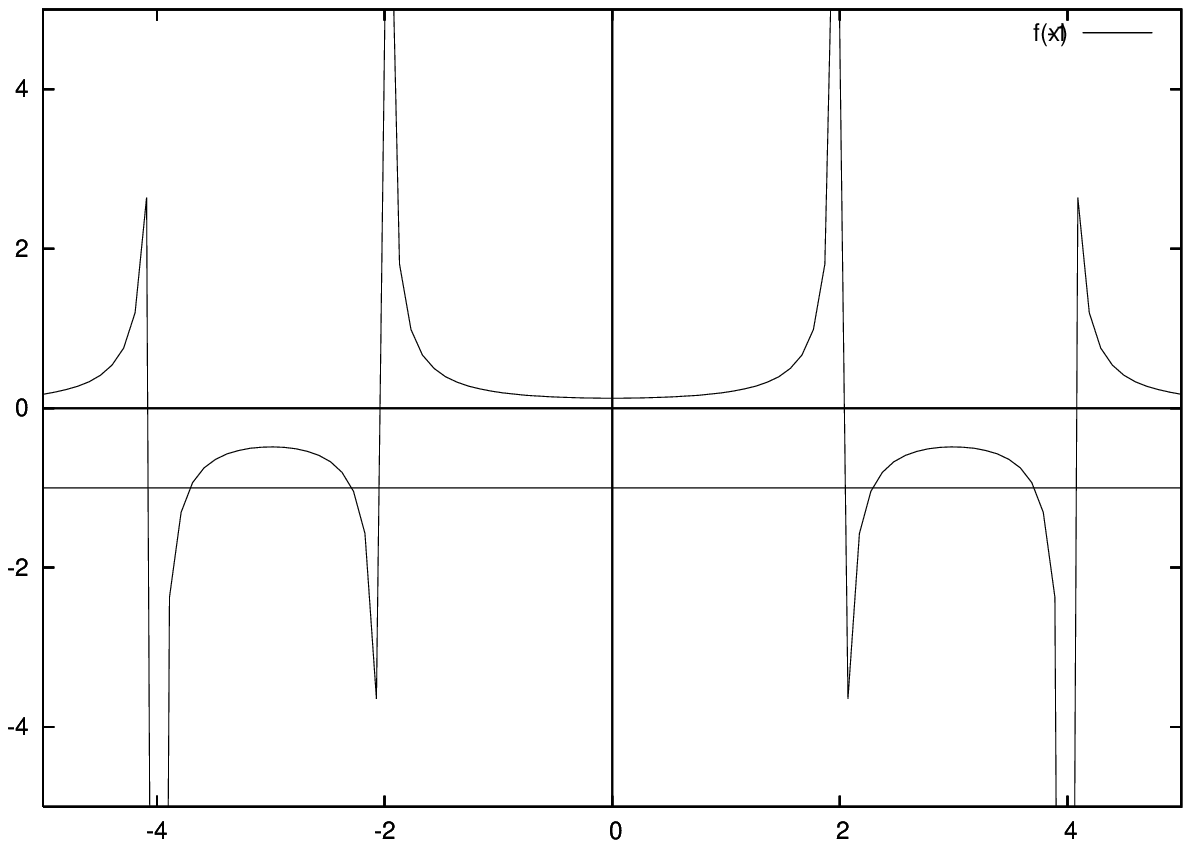,width=0.7\linewidth}\\
\bf{\small Figure 1: The left side of the dispersion 
equation against the wave frequency}
\end{center}

If the curves lying in the lower semiplane intersect
the straight line  $y = -1$, corresponding to $\varepsilon_0 = 0$,
the dispersion equation has four real solutions (i. e., such
a two-streamed system is stable in the linear approximation),
two of them corresponding to the positive, and two --- to
the negative energy of the wave.  (All the above holds true in 
the "supersonic" case, when $v_0 > c_s$ --- in the other case
the sketch on Fig. 1 would have looked differently.)

It is easy to show that there exists a range of parameters
in which such a system would be stable in the linear 
approximation.  In this case, a wave propagating in the
system can have either positive or negative sign in 
terms of energy.  The dispersion equation (\ref{dispersion doublecomponent}) 
can be reduced to biquadratic.  Introduce a nondismensional
constant $\eta$ in such a way that:
$$
v_0^2 = (1 + \eta)c_s^2
$$
The direct calculations show that the solutions  $\omega^2$ 
of the dispersion equation (\ref{dispersion doublecomponent})
are real when $\eta < 0$.  If $\eta > 0$, the solutions 
$\omega^2$ are real whenever
$$
\left[
\begin{array}{rcl}
0 \le k^2 c_s^2 \le \frac{\omega_j^2 \left(\sqrt{\eta + 1} - \sqrt{\eta}\right)}{2 \sqrt{\eta + 1}} \\
k^2 c_s^2 \ge \frac{\omega_j^2 \left(\sqrt{\eta + 1} + \sqrt{\eta}\right)}{2 \sqrt{\eta + 1}}
\end{array}
\right.
$$
and amount to
\begin{equation}
\label{frequency}
(\omega^2)_{1, 2} = k^2(2 + \eta)c_s^2 - \omega_j^2 \pm \sqrt{4(\eta + 1)k^4c_s^4 - 4(\eta + 1)k^2c_s^2\omega_j^2 + \omega_j^4}
\end{equation}

Note that in the particular case $k^2c_s^2 = \omega_j^2$
the solutions of the dispersion equation take the form:
$$
\begin{array}{rcl}
(\omega^2)_{1} &=& (2 + \eta)\omega_j^2 \\
(\omega^2)_{2} &=& \eta\omega_j^2 
\end{array}
$$
We see that, in this case, if $\eta > 0$, the dispersion
equation has four real solutions, two of them corresponding
to the positive values of the wave energy and two --- to
the negative ones.

Being interested in the solutions that are stable in the 
linear approximation (and thus corresponding to the ordinary
(longtitudinal) density waves), we should now find $k$
for which $\omega^2$ is nonnegative.  If $\eta > 0$, then
for $k$ satisfying the condition
\begin{equation}
\label{linear stability}
k^2c_s^2 \ge \omega_j^2 \frac{\sqrt{\eta + 1} +
\sqrt{\eta}}{2 \sqrt{\eta + 1}}
\end{equation}
all the four solutions $\omega$ are real, two of them
corresponding to the waves of the positive energy, and
two --- to the negative energy waves.

Checking the case $\eta = 0$ ($v_0 = c_s$), we see that
the system is stable in the linear approximation if 
$k^2c_s^2 \ge \omega_j^2$.  If $\eta < 0$, the system
demonstrates linear stabilities towards the perturbations
for which $k^2 c_s^2 \ge \frac{2 \omega_j}{|\eta|}$.
(One should remember that, when  $\eta < 0$,
$|\eta| \le 1$ by definition.)

\section{The Resonance Three, or Non-linear Interaction
Resulting in Explosion Instability}

Suppose that, being interested into all sorts of possible
perturbations of our two-streamed medium, we stick to the 
range of linear stability.  

We now consider plane waves as quasi-particles.  All
further calculations will be carried out in the
random phase approximation: we consider the waves 
(interpreted as quasiparticles) to be incoherent.
The quantities necessary for evaluation of the probability
of the event (resonance effect or any collision) are
provided by the second correction to the gravihydrodynamical
equations; the dispersion equation characterizing the
dependence of the frequency upon the wave vector we
have already obtained from the first correction,
or linear approximation.

Quasi-particle collisions proceed with the conservation
of energy and momentum.  The important question is, 
can those conditions be held for a process sketched
down on the scheme (Fig. 2) --- triple quasi-particle
production?
\begin{center}
\epsfig{file=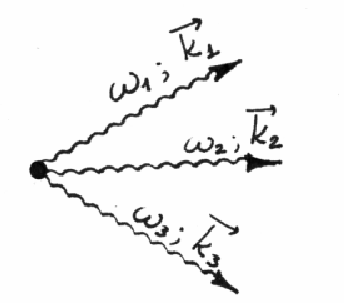,width=0.3\linewidth}\\
\bf{\small Fig. 2: Triple Quasi-particle Production}
\end{center}

The conservation laws require that
\begin{equation}
\label{conservation laws}
\begin{array}{rcl}
\omega_1 + \omega_2 + \omega_3 &=& 0 \\
\vec k_1 + \vec k_2 + \vec k_2 &=& 0
\end{array}
\end{equation}

As we shall see it below, such processes are indeed possible,
and can be found even in one-dimensional model.  Turn again
to the waves with $\mathbf k \| \mathbf v_0$.  As it can be
seen from the plot on Fig. 1, for each $k$, to the negative
energy wave corresponds the pair of $\omega$ having lesser 
absolute value.  Note by $\omega^{+}$ the solution of
the dispersion equation corresponding to the positive energy
of the wave, and by $\omega^{-}$ --- to the negative.
We are interested in the "resonance three", meaning 
simultaneous production of, say, one "positive" wave
and two "negative" ones.

\begin{equation}
\label{three waves}
\begin{array}{rcl}
\left(\omega_1^{-}\right)^2 &=& k_1^2 c_s^2 (2 + \eta) -
\omega_j^2 - \sqrt{4 (\eta + 1)k_1^4 c_s^4 - 4(\eta +
1)k_1^2c_s^2\omega_j^2 + \omega_j^4}\\
\left(\omega_2^{-}\right)^2 &=& k_2^2 c_s^2 (2 + \eta) -
\omega_j^2 - \sqrt{4 (\eta + 1)k_2^4 c_s^4 - 4(\eta +
1)k_2^2c_s^2\omega_j^2 + \omega_j^4}\\
\left(\omega_2^{+}\right)^2 &=& (k_1 - k_2)^2 c_s^2 (2 + \eta) -
\omega_j^2 + \\
&+&\sqrt{4 (\eta + 1)(k_1 - k_2)^4 c_s^4 - 4(\eta +
1)(k_1 - k_2)^2c_s^2\omega_j^2 + \omega_j^4}
\end{array}
\end{equation}
The first negative wave can be represented by $\omega_1^{-}, -k_1$,
the second wave of negative energy --- by $\omega_2^{-},
k_2$, and the wave of the positive energy, to complete
our "resonance three", will be represented by  $\omega^{+} = -
\omega_1^{-} - \omega_2^{-}$ and the corresponding wave
number $k_1 - k_2$.  (The dispersion law that we have here
is symmetric function with respect to $\omega$ (when $k$
is fixed) as well as with respect to $k$ with fixed
$\omega$.)
In order to prove that such a "resonance three" indeed
does exist, consider the case $k^2 c_s^2 >> \omega_j^2$.
In the second approximation with respect to
$\frac{\omega_j}{k c_s}$ we get:
\begin{equation}
\label{shortwave resonance accomplished}
\begin{array}{rcl}
k_1 = - \sqrt{\eta + 1} k&,& \omega_1 = \sqrt{\eta +
1}\left(\sqrt{\eta + 1} - 1\right)k c_s - \frac{\omega_j^2}{2 k c_s \sqrt{\eta + 1}}\\
k_2 = k&,& \omega_2 = \left(\sqrt{\eta + 1} -
1\right)k c_s - \frac{\omega_j^2}{2 k c_s \sqrt{\eta + 1}}\\
k_3 = \left(\sqrt{\eta + 1} - 1 \right) k&,&
\omega_3 = - \eta k c_s + \frac{\omega_j^2}{2 k
c_s}\frac{\sqrt{\eta + 1} + 1}{\eta \sqrt{\eta + 1}}
\end{array}
\end{equation}

Solving the equation $\omega_1 + \omega_2 + \omega_3 = 0$
in terms of $\eta$ gives the value $\eta = \frac{5}{4}$. 
For the short waves in the second approximation with
respect to a small parameter we have presented "in
quantities" the resonance three.  The relative velocity
of the two streams for the case under discussion must
be $2 v_0 = 3 c_s$.  Thus the triple quasi-particle
production, which can take place in a two-streamed
gravitating medium, is described by the conservation
laws (\ref{conservation laws}).

Note by $N_{k}$ the number of quasi-particles carrying
the 4-momentum\footnote{Here we have put $\hbar = 1$.}
$k$ (corresponding to the perturbations with the
wave vector $\mathbf k$ and the frequency
$\omega_{\mathbf k}$, $\mathbf k$ and $\omega_{\mathbf k}$
being related to each other by the dispersion equation).
Supposing $N_{ k} >> 1$ one can write:
\begin{equation}
\label{quasiparticles evolution}
\begin{split}
\frac{\partial N_{ k_1}}{\partial t}&= \sum_{ k_2} Q( k_1, 
 k_2) \left[ (N_{ k_1} + 1)(N_{ k_2} + 1)(N_{-  k_1 -  k_2} + 1) - N_{
 k_1}N_{ k_2}N_{-  k_1 -  k_2} \right ] \approx\\
&\approx \sum_{ k_2} Q( k_1,  k_2) \left [ N_{ k_1}N_{ k_2} +N_{ k
_2}N_{-  k_1 -  k_2} + N_{ k_1}N_{-  k_1 -  k_2} \right]
\end{split}
\end{equation}
Here $Q( k_1,  k_2)$ is the possibility of the triple 
quasi-particle production (or annihilation) of the
quasi-particles with the momenta  $( k_1,  k_2, -  k_1 - 
k_2)$.  As it is noted in \cite{Kingsep uchebnik}, after
the act of triple quasi-particle production is repeated
quite many times, the number of quasi-particles of the
three sorts that participate in the act can be considered
roughly the same.  Hence dismissing the indexes, we get:
$$
\frac{dN}{dt} = w N^2
$$
which leads us directly to the formula of the explosion
instability:
\begin{equation}
\label{explosion instability}
N(t) = \frac{N_0}{1 - N_0 w t}
\end{equation}
(Here $N$ is the number of quasi-particles of any of the
three resonant sorts dependent on time, $N_0$ --- the
initial number of the quasi-particles.  The coefficient
$w$ is to be interpreted as the possibility of the 
process, the value, inversely proportional to the
time of development of the instability.  In the next
section we shall calculate it for the one-dimensional
case.)

\section{Evaluation of the Time of Development of the
Explosion Instability in a Two-Streamed Gravitating
Medium}

The time $t_0 = (N_0 w)^{-1}$ of development of the
explosion instability (\ref{explosion instability})
is inversely proportional to the value $w$ --- the
possibility of the resonance process responsible for
this instability.  (Naturally, as one can see from
(\ref{explosion instability}), $w$ is the possibility
per $N$, the number of quasi-particles, which is 
itself a dimensional quantity.) 

We shall use the method developed by certain authors
for the decay instability in the random phases 
approximation in the beginning of 1960-ies.
(See, for instance, \cite{Kingsep uchebnik}.)
Since the gaseous galactic disk, with its particles
moving through the hypothetic halo of the dark matter,
is in the turbulent state, the random phase approximation
here seems adequate.

The linear dispersion equation was obtained in the
first approximation with respect to the amplitude;
considering the case of weak nonlinearity, one should
turn to the second approximation.

Consider the Fourier component of the Poisson equation
correcponding to the momentum $\mathbf k$ and 
frequency $\omega$.  For the sake of convenience
introduce the 4-dimensional wave vector $k = (\omega,
\mathbf k)$.  Whenever the inverse is not
clearly stated, speaking of the 4-vector k we would 
always have in mind $k = (\omega_{\mathbf k}, \mathbf k)$,
where $\omega_{\mathbf k} = \omega(\mathbf k)$ is the
solution of the dispersion equation 
$\varepsilon_0 (\omega, \mathbf k) = 0$.  By the above
relation a manifold in the 4-dimensional space is defined;
the 4-dimensional integration appearing in the further
calculations is performed over this manifold.
Below, among the indexes of the Fourier harmonics,
4-vector $k$ as well as the normal 3-vector $\mathbf k$,
will be used.
\begin{equation}
\label{Poisson}
-\mathbf k^2\varepsilon_0(\omega, \mathbf  k) \tilde \psi_{k} = 4 \pi G (\rho_{1 k}^{(2)} + \rho_{2 k}^{(2)})
\end{equation}
Here, as well as in the previous sections, $\varepsilon_0$
is the gravitational analogue of the dielectric
permettivity.  Index $(2)$ over the symbol of density
$\rho$ indicates the correction of the second approximation.
In the linear case $\varepsilon_0(\omega, \mathbf  k) = 0$
would have held; if we present the small nonlinear
correction to the frequency as $\delta \omega_N + i
\gamma_N$, then, as is easily seen, for our approximation
$$
\varepsilon_0(\omega, \mathbf k) \approx \left.\frac{\partial \varepsilon_0}{\partial \omega}\right|_{\omega = \omega_{\mathbf k}} (\delta \omega_N + i \gamma_N)
$$
Multiply both parts of the equality by some $\mathbf
k'$-component of the perturbed gravitational potential 
$\psi_{\mathbf k'}$ and then average it over phases:
$$
- (\delta \omega_N + i \gamma_N)\left.\frac{\partial \varepsilon_0}{\partial \omega}\right|_{\omega = \omega_{\mathbf k}} \mathbf k^2 I_{k} \delta (k + k') = 4 \pi G \left<\left(\rho_{1 k}^{(2)} +
\rho_{2 k}^{(2)}\right)\psi_{k'}\right>
$$
Here $I_{k}$ is the spectral intensity
($<\psi_{k}\psi_{k'}> =I_{k} \delta (k + k')$ 
in the random phases approximation).

The energy of the plane density wave is
$$
W_{k} = - \frac{1}{8 \pi G} \left.\frac{\partial (\varepsilon_0 \omega)}{\partial \omega}\right|_{\omega=\omega_{\mathbf k}}|g_{k}|^2
$$
Since $g_{k} = - (\nabla \psi)_{k} = - i \mathbf k
\psi_{k}$ and $\varepsilon_0(\omega_{\mathbf k}) = 0$,
for the number of quasi-particle with the 4-momentum 
$k$ one can write:
$$
N_{k} = \frac{W_{k}}{\omega_{\mathbf k}} = 
- \frac{1}{8 \pi G} \left.\frac{\partial \varepsilon_0}{\partial \omega}\right|_{\omega = \omega_{\mathbf k}} \mathbf k^2 I_{k}
$$
In the expression for the nonlinear frequency shift
$\delta \omega_N + i \gamma_N$ the real part can be
neglected compared to $\omega_{\mathbf k}$ --- it
corresponds to the renormalization of the linear
dispersion law.  The imaginary part $i \gamma_N$
corresponds to the nonlinear spectre evolution.
Note that for $\psi\sim e^{-i (\omega + i \gamma_N)}$
the number of quasi-particles will be $N_{k} \sim e^{2
\gamma_N t}$, so that multiplification by $\gamma_N$
for the function $ N_{k}(t) $ is equivalent to the
action of the operator of the differentiation over the
time: $\gamma_N \sim \frac{1}{2}\frac{\partial}{\partial t}$.
Taking all this into account, note that the equation (\ref{Poisson})
transforms into:
\begin{equation}
\label{quasiparticles evolution 2}
\frac{\partial N_{ k}}{\partial t}\delta(k + k') = - i \left<\left(\rho_{1 k}^{(2)} + \rho_{2 k}^{(2)}\right)\psi_{k'} \right>
\end{equation}
On the other hand, the structure of the time evolution
of the number of quasi-particles is determined by the 
equation  (\ref{quasiparticles evolution}), or, in the
integral form,
\begin{equation}
\label{quasiparticles evolution 3}
\frac{\partial N_{\mathbf k}}{\partial t} = \int w(\mathbf k, \mathbf q)\delta(\omega_{\mathbf k} + \omega_{\mathbf q} + \omega_{- \mathbf k - \mathbf q})\left(N_{\mathbf q}N_{- \mathbf k - \mathbf q} + N_{\mathbf k}N_{\mathbf q} + N_{\mathbf k}N_{- \mathbf k -\mathbf q}\right)dq
\end{equation}

To make the long calculations short, we now set to follow
the evolution of only one term of the sum under the symbol
of integration, say, $N_{\mathbf q}N_{- \mathbf k - \mathbf q}$.
We shall sort of exstract it from the second correction to
the density; the analogous expressions for the rest two terms
will be easy to construct; besides, the coefficient (i. e.
the possibility in which we are interested) preceding them
in the expressions like \ref{quasiparticles evolution 3}
is the same.  We take the second correction to the 
density in the following form:
\begin{equation}
\label{density}
\rho_{\alpha k} = \int \nu_{\mathbf k, \mathbf q}^{\alpha}\left(\psi_{q} \psi_{- k - q} - \left< \psi_{q} \psi_{- k - q} \right> \right) d q
\end{equation}
Here $\alpha = 1,2$ is the number of the corresponding component
of the medium, $\nu_{\mathbf k, \mathbf q}^{\alpha}$ is the
coefficient to be calculated a few passages later, from 
the Euler equations.  The 3-dimensional indexes $\mathbf k, \mathbf q$
are used instead of the 4-dimensional ones because three components
of the wave vector are sufficient to fully define the coefficient
$\nu$.  The average value $\left<\psi_{q}\psi_{- k - q}\right>$
corresponds to the evolution of the phone and should be
subtracted.

Substituting (\ref{density}) into (\ref{quasiparticles evolution 2}),
we get:
\begin{equation}
\label{d}
\frac{\partial N_{k}}{\partial t}\delta(k + k') = -i \int \left(\sum_{\alpha = 1, 2} \nu_{\mathbf k, \mathbf q}^{\alpha}\right) \left<\psi_{k'} \left(\psi_q\psi_{-k - q} - <\psi_{q}\psi_{- k - q}>\right)\right> d q 
\end{equation}

Substitute now the Fourier-component $\psi_{k'}$ by
its nonlinear expansion (through the density from
(\ref{Poisson})):
$$
\psi_{k'}^{(2)} = - \frac{4 \pi G}{k'^2 \varepsilon_0(\omega', \mathbf k')} \int \left(\sum_{\alpha}\nu_{\mathbf k', \mathbf q'}^{\alpha}\right) \left(\psi_{q'}\psi_{- k' - q'} - <\psi_{q'}\psi_{- k' - q'}>\right) d q'
$$ 
in the course of this substitution we would have to 
perform the averaging procedure.  The randomness of the
phases results in reducing of all the correlations to
binary, so that:
\begin{equation*}
\begin{split}
\left< \psi_{q'}\psi_{- k' - q'} \psi_{q} \psi_{- k - q}\right> &= <\psi_{q'}\psi_{q}><\psi_{-k' - q'}\psi_{- k - q}> + <\psi_{q'}\psi_{- k' - q'}><\psi_{q}\psi_{- k - q}> +\\
&+ <\psi_{q'}\psi_{- k - q}><\psi_{- k' - q'} \psi_{q}> 
\end{split}
\end{equation*}
The middle term here vanishes by the virtue of (\ref{d}), 
and the rest two terms are symmetrical with respect to 
replacement $q' \to - k' - q'$, so that one of them can
be doubled and the other one written off:
\begin{equation*}
\begin{split}
\left< \psi_{q'}\psi_{- k' - q'} \psi_{q} \psi_{- k - q}\right> &= 2 I_{q}I_{- k - q} \delta(q + q') \delta(k + k') = \\
&= 2\cdot (8 \pi G)^2 N_{q} N_{- k - q} \left[ (\mathbf k + \mathbf q)^2  \left.\frac{\partial \varepsilon_0}{\partial \omega}\right|_{\omega = \omega_{-\mathbf k - \mathbf q}} \mathbf q^2 \left.\frac{\partial \varepsilon_0}{\partial \omega}\right|_{\omega = \omega_{\mathbf q}}\right]^{-1}\times \\
&\times\delta(q + q') \delta (k + k')
\end{split}
\end{equation*}

Using an approximate equality in terms of generalized
functions:
$$
\varepsilon_0^{-1}(\omega'_{\mathbf k'}, \mathbf k') \simeq \left[ (\omega' - \omega'_{\mathbf k'})\left.\frac{\partial \varepsilon_0}{\partial \omega}\right|_{\omega = \omega'_{\mathbf k'}}\right]^{-1} \simeq \left(\left.\frac{\partial \varepsilon_0}{\partial \omega}\right|_{\omega = \omega'_{\mathbf k'}}\right)^{-1} i \pi \delta(\omega' - \omega'_{\mathbf k'})
$$
and substituting all that into (\ref{d}), performing
integration over $d q'$ and $d k'$, for the evolution
of the number of quasi-particles (tracing the 
transformations of one of the first two terms of
(\ref{quasiparticles evolution 3}) --- $N_{\mathbf q} N_{- \mathbf k - \mathbf q}$ --- we end up with the following expression:
\begin{equation}
\label{quasiparticles evolution almost done}
\begin{split}
\frac{\partial N_{k}}{\partial t}  &= i \tau \int d q (8 \pi G)^3 (\sum \nu_{\mathbf k, \mathbf q}^{\alpha})^2 \times \\
&\times \left[\mathbf k^2 \left.\frac{\partial \varepsilon_0}{\partial \omega}\right|_{\omega = \omega_{\mathbf k}} (\mathbf 
k + \mathbf q)^2 \left.\frac{\partial \varepsilon_0}{\partial
\omega}\right|_{\omega = \omega_{-\mathbf k - \mathbf q}} \mathbf q^2
\left.\frac{\partial \varepsilon_0}{\partial
\omega}\right|_{\omega = \omega_{\mathbf q}} \right]^{-1}
\times \\
&\times N_{q} N_{- k - q} 
\end{split}
\end{equation}

Here $\tau$ is the dimensional coefficient, by the order of
magnitude similar to $L/v_0$; $\tau = \frac{V}{(2 \pi)^3} \int \delta(\omega' - \omega_{\mathbf k'}) d \mathbf k'$.  Thus we see that the 
expression for the probability has the following structure:
\begin{equation}
\label{probability}
w = -(8 \pi G)^3\pi  \tau \left(\sum_{\alpha}\nu_{\mathbf k, \mathbf q}^{\alpha}\right)^2\left[\mathbf k^2\left.\frac{\partial \varepsilon_0}{\partial \omega}\right|_{\omega = \omega_{\mathbf k}} (-\mathbf k - \mathbf q)^2\left.\frac{\partial \varepsilon_0}{\partial \omega}\right|_{\omega = \omega_{-\mathbf k -\mathbf q}} \mathbf q^2 \left.\frac{\partial \varepsilon_0}{\partial \omega}\right|_{\omega = \omega_{\mathbf q}}\right]^{-1}
\end{equation}

Now one has to determine the coefficient $\nu_{\mathbf k, \mathbf q}^{\alpha}$
given by relation (\ref{density}).  The second correction to
the continuity equation is:
\begin{equation}
\label{continuity (2)}
\frac{\partial \rho_{\alpha}^{(2)}}{\partial t} + \rho_{\alpha}^{(1)} div \mathbf v_{\alpha}^{(1)} + (\mathbf v_{\alpha}^{(1)}\nabla)\rho_{\alpha}^{(1)} + \rho_{0 \alpha}div \mathbf v_{\alpha}^{(2)} + (\mathbf v_{0 \alpha} \nabla)\rho_{\alpha}^{(2)} = 0
\end{equation}

The Euler equation in the second approximation:
\begin{equation}
\label{Euler (2)}
\frac{\partial \mathbf v_{\alpha}^{(2)}}{\partial t} = -
(\mathbf v_{\alpha}^{(1)} \nabla) \mathbf v_{\alpha}^{(1)} -
(\mathbf v_{\alpha}^{(0)} \nabla) \mathbf v_{\alpha}^{(2)}
\end{equation}

Fourier-expansion for  $(\omega, \mathbf k)$-component of
the continuity equation gives:
\begin{equation}
\omega \rho_{\alpha, \mathbf k}^{(2)} = \int d \mathbf q
\rho_{\alpha \mathbf q}^{(1)}\left( (\mathbf k - \mathbf q), \mathbf 
v_{\alpha, \mathbf k - \mathbf q}^{(1)}\right) + \int d \mathbf q
\left(\mathbf q, \mathbf v_{\alpha, \mathbf k - \mathbf 
q}^{(1)}\right) \rho_{\alpha, \mathbf q}^{(1)} + \rho_{0
\alpha}\left(\mathbf k \mathbf v_{\alpha, \mathbf k}^{(2)}\right) +
\rho_{\alpha, \mathbf k}^{(2)}(\mathbf k \mathbf{v_{0 \alpha}})
\end{equation}

Rewriting the same relation in more symmetrical form:
\begin{equation}
\label{continuity Fourrier}
\rho_{\alpha, \mathbf k}^{(2)}(\omega - \mathbf k \mathbf v_{0
\alpha}) = \frac{1}{2} \int d \mathbf q \left[\rho_{\alpha
\mathbf q}^{(1)}(\mathbf k, \mathbf v_{\alpha, \mathbf k - \mathbf 
q}^{(1)}) + \rho_{\alpha, \mathbf k - \mathbf q}^{(1)}(\mathbf k,
\mathbf v_{\alpha, \mathbf q}^{(1)})\right] + \rho_{0
\alpha}(\mathbf k, \mathbf v_{\alpha, \mathbf k}^{(2)})
\end{equation}

The values of the linear approximation are related in
the following way:
$$
\rho_{\alpha, \mathbf k}^{(1)} = \rho_{\alpha 0}\frac{\mathbf k
\mathbf v_{\alpha, \mathbf k}^{(1)}}{\omega - \mathbf k \mathbf v_{0
\alpha}}
$$

Fourier transform applied to the Euler equations
gives:
\begin{equation}
\label{Euler Fourrier}
(\omega - \mathbf k \mathbf v_{0 \alpha}) \mathbf v_{\alpha \mathbf 
k}^{(2)} = \frac{1}{2}\int d \mathbf q \left[\mathbf v_{\alpha
\mathbf q}^{(1)} \left(\mathbf q, \mathbf v_{\alpha, \mathbf k - \mathbf 
q}^{(1)} \right) + \mathbf v_{\mathbf k - \mathbf 
q}^{(1)}\left(\mathbf k - \mathbf q, \mathbf v_{\mathbf 
q}^{(1)}\right)\right]
\end{equation}

In order to obtain the result in the analitical form,
we now turn to 1-D case.  As in previous sections,
let $\rho_{01} = \rho_{02} = \rho_{0}$; $v_{01} = - v_{02} = v_0$.
Below the index <<$k$>>, marking the Fourier component,
will mean now the one-dimensional "wave number", not the
4-momentum.

From the Fourier transform of the linear approximation
one easily obtains:
$$
v_{1,k}^{(1)} = \frac{k (\omega_{k} -
kv_0)}{(\omega_{k} - k v_0)^2 - k^2 c_s^2}\psi_k
$$
$$
\rho_{1,k}^{(1)} = \frac{k^2 \rho_0}{(\omega_{k} - kv_0)^2
- k^2c_s^2}\psi_k
$$
Changing $v_0 \to - v_0$ and $(1)$ to $(2)$ in the 
above expression, we get the similar relation for
the second component.

Substituting (\ref{Euler Fourrier}) into (\ref{continuity
Fourrier}) and taking into account the above expressions
for the Fourier components of the velocity and density
in the linear approximations as related to the 
corresponding Fourier component of the amplitude of
the gravitational potential, also having in mind that 
$\nu_{-k,q}^1 = \nu_{k,q}^2$ due to the symmetry of the
initial conditions, we obtain the coefficient:
\begin{equation}
\label{density to potential}
\nu_{k,q}^2 = \frac{1}{2} \rho_0 \times \frac{k q (k + q)}{A_q^{-}
A_{-k-q}^{-} (\Omega_{-k}^{-})^{2}} \times T_{1}^{2,3}(-)
\end{equation}

Here (in the above) for each $\chi = k, q, - k - q$
the following notations have been introduced:
$$
\Omega_{\chi}^{\pm} = \omega_{\chi} {\pm} \chi v_0
$$
$$
A_{\chi}^{\pm} = (\Omega_{\chi}^{\pm})^2 - \chi^2 c_s^2
$$
and
$$
T_1^{2,3}(\pm) = q\Omega_{-k-q}^{\pm}\Omega_{-k}^{\pm} + (-k - q)\Omega_q^{\pm}\Omega_{-k}^{\pm} + (-k)\Omega_q^{\pm}\Omega_{-k-q}^{\pm}
$$

For the first component of the medium (the stream moving
with the positive velocity $v_0$) the corresponding 
coefficient $\nu_{k,q}^1$ can be obtained from  $\nu_{k,q}^2$
by replacement $v_0 \to -v_0$.  In the above passages,
performing the Fourier transform, we have not usually
mentioned the dimensional coefficient; it is easy to
see that the latter will be important only in calculation
of the coefficient $\tau$, which in one-dimensional model
is equal to:
$$\tau = \frac{L}{2 \pi} \int \delta (\omega' - \omega_{k'}) dk' =\frac{L}{2 \pi} \left. \frac{dk'}{d \omega'}\right|_{\omega' = \omega_{k'}} $$
$L$ being the characteristic scale of the problem, the
linear size of the system in the direction parallel
to the velocity of both streams.  Calculating this
expression for $k' = k$, $\omega_{k'} = \omega_k$,
we get:
$$ \tau = \frac{L}{2 \pi} \times \frac{(A_k^{+})^2 \Omega_{-} + (A_k^{-})^2 \Omega_{+}}{(-\omega v_0 + k \eta c_s^2)(A_k^{+})^2 + (\omega v_0 + k \eta c_s^2)(A_k^{-})^2} $$
Thus the probability is, effectively, found.  The expression
for the probability even in one-dimensional case looks
somewhat awkward:
\begin{equation}
\label{probability_2}
w = \frac{\pi \tau}{4 \rho_0}\times
\frac{\left(T_1^{2,3}(-)A_q^{+}A_{-k-q}^{+}(\Omega_{-k}^{+})^2
+
T_1^{2,3}(+)A_q^{-}A_{-k-q}^{-}(\Omega_{-k}^{-})^2\right)^2(A_k^{-})^2(A_{k}^{+})^2}{(\Omega_k^{+})^4(\Omega_k^{-})^4
D_k D_q D_{-k-q}}
\end{equation}
Here for all $\chi = k, q, - k - q$ the following notation
is introduced:
$$
D_{\chi} = \Omega_{\chi}^{-}(A_{\chi}^{+})^2
+ \Omega_{\chi}^{+} (A_{\chi}^{-})^2
$$
The expression obtained above, in principle, provides 
a possibility to evaluate the time of development of
the instability, if the hydrodynamical parameters of
the problem (i. e. density, speed of sound) are known
and the "resonance three" are determined.  In our case,
since the two streams going through each other in 
the opposite directions are hydrodynamically
identical, and the "resonance three" has been obtained,
all we have to know to evaluate the time of development
of the explosion instability is the density and the 
speed of sound.

For the resonance three determined in one of the 
previous sections, if the characteristic size $L$
is taken to be the jeans size of one of the components,
then the value $w$, calculated according to (\ref{probability_2}),
acquires the following structure:
$$
w = \frac{4 \pi \omega_j^2}{\rho_0 c_s^2 \beta}
$$
$\beta$ being the relation of the jeans wavelength 
to the wavelength of the perturbation in question.
Supposing that in gravihydrodynamics, like in plasma
physics, the degree of the turbulence can be
evaluated by the parameter $\frac{W}{nT} << 1$, where
$W$ is the density of the noise energy, we get the
evaluation for the time of development of the 
explosion instability:
$$
t_{ins} \sim \alpha(k) \frac{\rho_0 c_s^2}{n_0 T \omega_j}
$$
$\alpha(k)$ being a dimensionless coefficient which is the
less the greater the turbulence degree of the medium 
$\frac{W}{nT}$ is.  If $W(k)$ has a pointed peak then
for the value of $k$ corresponding to that maximum 
$\alpha(k) << 1$.  Since, obviously, the second factor
in the expression for  $t_{ins}$ has the same order of
magnitude as the inverse jeans frequency (say, $10^{15} -
10^{16}$ s for the interstellar gas), that would mean
that, for those $k$, the instability would be able to
develop.  It is important to have in mind that what we
deal with is a nonlinear process, achieving the 
infinite amplitude in finite times (ideally speaking).
Comparing linear exponential instabilities we take
the definitely fastest one in terms of increments;
to say that one of the instabilities is fast enough
to develop against the background of the other, we
require a strong inequality $\gamma_1 >> \gamma_2$
($\gamma_1$, $\gamma_2$ being the increments).  But when
one of the instabilities in question is explosion 
instability --- the simple inequality is enough to
conclude that it is faster\footnote{In our case the
actual characteristic sizes are not, in fact, jeans 
wavelengths: they are determined by the conditions
limiting the applicability of our plane approximation.
In the real cosmic space geometry, however, the
large-scale instabilities are stabilized by rotation.}.

Returning to the real astronomical situation, it should
be noted that the matter of the gaseous disk and the
dark matter can hardly be considered hydronamically
identical.  To take the hydrodynamical difference 
between the two streams into account should not 
change dramatically the expression for the probability:
each coefficient $\nu_{k, q}^{\alpha}$, $\alpha = 1,2$,
will be expressed through the equilibrium density of 
the corresponding component, and the auxiliary
constructions like $D_{\chi}^{\pm}$ would get additional
indexes, so that:
$$
A_{\chi}^{+} = (\omega_{\chi} + \chi v_0)^2 - k^2 c_{s1}^2
$$
$$
A_{\chi}^{-} = (\omega_{\chi} - \chi v_0)^2 - k^2 c_{s2}^2
$$                                                                       
Yet the dispersion equation would not be analitically
solvable any more; numerical calculations would be
required.

\section{Conclusions}

In the present work, we have obtained the expression for the
density wave in a gravitating medium, analogous to the
corresponding formula in plasma physics.  The analogy has
been toyed with by the theoretists of both fields, but
seems to have never been clearly formulated and
demonstrated in the general case.  The important role
in this parallel between the two regions of physics
is played by a physical value describing the responce of
the distribution of the density of the gravitating matter
to a change in the gravitational field.  In electrostatics
and electrodynamics the analogous value is the dielectric
permettivity of the medium.

Having thus legitimized certain analogies between the plasma
physics and physics of gravitating media that have been
noted before, we concentrated on development of one
of those analogies, related to the possibility of development 
of the explosion instability in a two-streamed gravitating
medium.

We have shown that, just as in a two-streamed plasma,
in the two-streamed gravitating medium simultaneous
propagation of the waves of positive and negative
energies is possible.  On the example of two streams
moving through each other in opposite directions
and having the same hydrodynamic characteristics
it was demonstrated that a multicomponent gravitating
medium can provide the background for development of
the explosion instability.

The explosion instability is faster than the linear,
exponential instabilities; the amplitude of the 
perturbation reaches infinity within finite time.
Using the standard method developed in plasma
physics, we have demonstrated the way to calculate
this time for one-dimensional model.

Stabilization of the explosion instability is achieved
through restructuralization of the dispersion equation.
The dependence of the frequency of the perturbation on
the wave vector that we have used in our calculations
was obtained from the linear approximation.  For large
enough amplitudes, this dependence changes its form,
the "resonance three" do not resonate any more, so that
triple production of these quasi-particles becomes
impossible.

Classification of various scenarios, composition
of a librarian's catalogue of all possible instabilities
has been long recognised as an important task in plasma
physics.  In graviphysics, the instabilities responsible
for the most part of the observed phenomena and structures
in the Universe probably deserve no less attention on
the part of physicists and astronomers.

The authors would like to express their gratitude to
A. L. Barabanov, K. V. Chukbar and A. M. Fridman
for fruitful discussions.

\end{document}